\documentclass[10p,final,journal]{IEEEtran}

\usepackage{amssymb,amsmath,color,graphicx, amsbsy}
\usepackage{graphicx}
\usepackage{subfigure}
\usepackage{cite}
\usepackage{algorithm}
\usepackage{algorithmic}
\usepackage{amsmath}
\usepackage{amsfonts}
\usepackage{mathtools}
\usepackage{amssymb}
\usepackage{graphicx}
\usepackage{verbatim}
\usepackage{epstopdf}
\usepackage{lettrine}
\usepackage{enumitem}
\usepackage{caption}

\usepackage{cleveref}
\usepackage[export]{adjustbox}
\usepackage[T1]{fontenc}
\usepackage[utf8]{inputenc}
\usepackage{authblk}
\usepackage[normalem]{ulem}
\usepackage{mwe}
\usepackage{xcolor}

\DeclareMathOperator{\EX}{\mathbb{E}}

\begin{document}

\title{Event Detection in Micro-PMU Data: A Generative Adversarial Network Scoring Method}

\author[1]{Armin Aligholian\thanks{This work is supported by UCOP  LFR-18-548175 and DoE  EE 8001 grants. The corresponding author is H. Mohsenian-Rad, e-mail: hamed@ece.ucr.edu.}}

\author[1]{{Alireza Shahsavari}}

\author[2]{Ed Cortez}
\author[3]{Emma Stewart}

\author[1]{Hamed Mohsenian-Rad}

\affil[1]{Department of Electrical and Computer Engineering, University of California, Riverside, CA, USA}
\color{black}

\affil[2]{Energy Delivery Engineering Group, Riverside Public Utilities, Riverside, CA, USA}

\affil[3]{Defense Infrastructure Group, Lawrence Livermore National Laboratory, Livermore, CA, USA \vspace{-0.7cm}}

\renewcommand\Authands{ and }

\maketitle

\thispagestyle{empty}

\pagestyle{empty}

\begin{abstract}

A new data-driven method is proposed to detect events in the data streams from \emph{distribution-level phasor measurement units}, a.k.a., micro-PMUs. The proposed method is developed by constructing \emph{unsupervised deep learning} anomaly detection models; thus, providing event detection algorithms that require \emph{no or minimal human knowledge}. First, we develop the core components of our approach based on a Generative Adversarial Network (GAN) model. We refer to this method as the \emph{basic} method. It uses the same features that are often used in the literature to detect events in micro-PMU data. Next, we propose a second method, which we refer to as the \emph{enhanced} method, which is enforced with additional feature analysis. Both methods can detect \emph{point signatures} on single features and also \emph{group signatures} on multiple features. This capability can address the unbalanced nature of power distribution circuits.
The proposed methods are evaluated using \emph{real-world} micro-PMU data. We show that both methods highly outperform a state-of-the-art statistical method in terms of the event detection accuracy. The enhanced method also outperforms the basic method. 

\vspace{0.27cm}


\end{abstract}

\vspace{-0.2cm}

\section{Introduction} \label{sec:introduction}

The voltage and current measurements that are reported by distribution-level phasor measurement units, a.k.a, micro-PMUs, have high-resolution and high-precision. They significantly enhance our visibility into the distribution grid, \cite{emma, hamedreview}.  
%
%
Applications of micro-PMU data include topology and phase identification \cite{farajtopology}, load modeling, \cite{shahload}, state estimation \cite{farajstate}, asset monitoring \cite{shahcap}, and distribution system cyber-security \cite{jameicyber}. 


An important and emerging class of studies when it comes to micro-PMU data is to investigate the \emph{events} in distribution systems. Here, an event is defined rather broadly and may refer to load switching, capacitor bank switching, connection or disconnection of distributed energy resources (DERs), inverter malfunction, a minor fault, a signature for an incipient fault, etc. \cite{ref2, ref4, ref9}. {Undoubtedly,} the very first step to investigate the events in micro-PMU data is to \emph{detect} them from the large volume of data that is being collected. Note that, each micro-PMU reports over \emph{one hundred million data points} every day.  

The literature on event detection in micro-PMU data can be divided into two broad classes; namely \emph{statistical methods}, such as in \cite{ref5, shahresidual,Ardakanian}, and \emph{machine learning methods}, such as in \cite{bundle,ref7}. Such common categorization have been utilized in other areas of study within smart grid literature, for instance anomaly detection in smart meters \cite{Armin} and IoT systems \cite{Ehsan}. The statistical method in \cite{ref9} uses the absolute deviation around median, combined with dynamic window sizes. In \cite{Ardakanian}, the analysis of the inverse power flow problem is combined with the turning point test method to detect events. In \cite{ref5}, the physical equations of the power distribution circuits are combined with techniques from statistical quality control in order to develop a hierarchical anomaly detection architecture that uses data from optimally placed micro-PMUs.

While we use the state-of-the-art statistical method in \cite{ref9} as a \emph{benchmark} for performance comparison in this paper, our approach here is rather based on machine learning. In \cite{bundle}, 
a machine learning method, called ensembles of bundle classifiers, is used to train multiple classifiers based on multiple instances of the same predetermined event, so that the patterns of that event are learned in order to detect more instances of that event in the micro-PMU data. In \cite{ref7}, a hidden structure semi-supervised machine learning model is established to combine micro-PMU data for both labeled and unlabeled events. A parametric dual optimization procedure is used to tackle the non-convex learning objective function. 

The event detection method in \cite{bundle} is based on \emph{supervised} machine learning. Also, the method in \cite{ref7} is based on \emph{semi-supervised} machine learning. In both cases, full or partial  \emph{expert knowledge} is needed in order to establish the event detection tool. In contrast, in this paper, we develop an \emph{unsupervised} method to detect events in micro-PMU data. This eliminates the need for human labor; which makes our proposed approach suitable for implementation in practice when we work with large volumes of micro-PMU data. It is worth adding that unsupervised learning is previously used in the analysis of micro-PMU data, but as a feature extraction tool for \emph{event classification} \cite{livani}. {Therefore}, it has \emph{not} been used for event detection, which is the focus of this paper.

The event detection methods that are proposed in this paper, work by constructing deep learning Generative Adversarial Network (GAN) models. 
%
%
%
%
The contributions are as follows: 






\begin{itemize}


\item To the best of our knowledge, this is the first paper to develop event detection methods for micro-PMU data based on GAN models. Two such methods are proposed. The first one, called the basic method, trains a single GAN model. The second one, called the enhanced method, involves additional analysis of the features of the micro-PMU data; which leads to training two GAN models. 


\item Both methods are \emph{unsupervised} deep learning methods, which require no or minimal human knowledge; which makes them suitable for automated and scalable operation. Furthermore, they can detect both \emph{point-signatures} and \emph{group-signatures} in micro-PMU data. This is an important capability because of the unbalanced nature of power distribution circuits; where many events may affect only a subset of the features on only one or two phases. 


\item Real-world micro-PMU data is used to evaluate the proposed event detection methods. In order to create a reference, first, {more than} 1000 events of different kinds are extracted manually from the micro-PMU data within a given period of time. It is observed that both the basic and the enhanced methods highly outperform a prevalent statistical method. 
The advantage is particularly major for the events that cause \emph{small changes in magnitude}. Also, the enhanced method outperforms the basic method; thus confirming the importance of the revised  model. 

    
    
    
\end{itemize}

 







\section{Methodology} \label{Methodology}



\subsection{Basic Method} \label{Genral GAN}


In its core, the proposed basic event detection method uses a GAN model which has two components, a \emph{generator} and a \emph{discriminator}. The generator is a deep neural network that tends to \emph{produce} data samples that
follow the distribution of the historical training data. The discriminator is a deep neural network that tends to \emph{distinguish} between the data samples generated by the generator and the true historical data. By training the generator and the discriminator subsequently and iteratively, the GAN model can achieve an \emph{equilibrium}, at which the discriminator can no longer distinguish between the distribution of the generated samples and the historical data. 

\vspace{0.05cm}

\subsubsection{Features}

As in \cite{ref9, ref5}, we use the following time-series as the features to train the GAN model in our basic method:  1) magnitude of voltage, i.e., $V$; 2) magnitude of current, i.e., $I$; 3) active power, i.e., $P$; and 4) reactive power, i.e., $Q$. All these features are defined separately for each three phases. Therefore, in total, the GAN model is trained with 12 time-series. Note that, while micro-PMUs measure $V$ and $I$ directly, $P$ and $Q$ are obtained rather indirectly by combining $V$ and $I$ with the measurements on voltage phase angle and current phase angle, which are both provided by micro-PMUs. 

\vspace{0.05cm}

\subsubsection{Generator}

    
It takes a noise vector $z$ from a distribution function $p_z(z)$, such as $z \sim \mathcal{N}(\mu_z , \sigma_z^{2}$), and {tries to produce samples} similar to the ones from the true sample distribution. 
We seek to train a neural network $G(z,\theta_g)$ to generate samples {which} follow the distribution of the historical data. Here, $\theta_g$ denotes weights of the generator network.
%
Mathematically, we seek to minimize the following objective function \cite{ref13}: 
\begin{equation}
        \begin{multlined}
           \frac{1}{N} \sum_{i=1}^{N} \big[log(1-D(G(z_i))) \big],
        \end{multlined}
        \label{gengoal}
\end{equation}
where $N$ is the number of samples in each training batch, $D$ is the discriminator function, $G$ is the generator function, {and $z_i$ is the random vector for $i$th generated sample}. 
In order to train the generator, after forward propagation, we need to update the generator parameters by calculating gradient and using a proper optimizer, such as Adam optimizer \cite{adam}. 
%

\vspace{0.05cm}

\subsubsection{Discriminator}

It is meant to distinguish between the fake data samples generated by the generator and the real measurements. Our goal is to train a neural network $D(x, \theta_d)$, which creates a single scalar value as its output. Here, $x$ is the vector of the actual measurement data and $\theta_d$ is the weights of the discriminator network. The primary objective of the discriminator is to maximize the probability of distinguishing between the true measurement data and the data generated by the generator. Therefore, we seek to minimize: 
%
%
\begin{equation}
        \begin{multlined}
             \frac{1}{N} \sum_{i=1}^{N} \big[ log(D(x_i)) + log(1-D(G(z_i))) \big],
        \end{multlined}
        \label{disgoal}
\end{equation}
where {$x_i$ is the $i$th real sample} and the second term is the same as the term in (\ref{gengoal}). 
%
%

Together, the generator and the discriminator play a \textit{min-max} game with the following value function: 
    \begin{equation}
        \begin{multlined}
            V(G, D) = \EX_x\sim p_{data}(x)[log(D(x))] \:  +  \; \: \\
              \EX_x \sim p_{z}(z)[log(1-D(G(z)))].
        \end{multlined}
        \label{eq1}
    \end{equation}


\begin{algorithm}[t]
    \caption{Event Detection - Basic Method}
  \begin{algorithmic}[t]
  
    \STATE \textbf{Input:} Training data and test data: $V$, $I$, $P$ and $Q$. 
    
    \STATE \textbf{Output:} Event Detection Flag $F$.
    
    \STATE \textbf{// Learning Phase}

    \STATE Train the $GAN$ model.
    
    \STATE Use the Discriminator as scoring function $D^*(\cdot)$.

    \STATE Calculate the scores for the training data.

    
    \STATE Fit a Normal PDF $\mathcal{N}(\mu,\sigma^{2})$ to the obtained scores. 
    
    
    \STATE \textbf{// Detection Phase}
    
    \STATE \textbf{For} each new micro-PMU test data \textbf{Do}
    
    \STATE $ \ \ \ $ Calculate the score $s$ using  $D^*(\cdot)$.
    
    \STATE $ \ \ \ $ \textbf{If} $s \notin (\mu - z_p \delta, \mu + z_p \delta)$ \textbf{Then}
    
    \STATE $ \ \ \ \ \ \ $ $F = 1$ // Event
    
    \STATE $ \ \ \ $ \textbf{Else} 
    
    \STATE $ \ \ \ \ \ \ $ $F = 0$ // No Event
    
    \STATE $ \ \ \ $ \textbf{End} 
    
    \STATE \textbf{End}
    
    
    \label{alg_event_scoring_basic}
  \end{algorithmic}
\end{algorithm}

\subsubsection{Training}

Both the generator and discriminator are formed with Long Short-Term Memory (LSTM) modules, which 
are connected back-to-back to capture the relationship between different features 
and their time dependencies. The micro-PMU data is normalized and segregated into sequences of training blocks. 
%
%
%
%
The value of $V(G,D)$ can attain its global {optimum} by satisfying the following two conditions:
\begin{itemize}
    \item \textbf{C1:} For any fixed $G$, the optimal discriminator {$D^*$} is:
    \begin{equation}
        D_G^*(x)=\frac{p_{data}(x)}{p_{data}(x)+p_{g}(x)}.
        \label{eq2}
    \end{equation}
    \item \textbf{C2:} There exists a global solution such that: 
    \begin{equation}
        \begin{aligned}
             \min(\mathrel{\mathop{\max_{D}(V(G,D))}}) \Longleftrightarrow p_{g}(x)=p_{data}(x).
        \end{aligned}
        \label{eq3}
    \end{equation}
\end{itemize}

If these conditions are not satisfied at the equilibrium, then the training is repeated with new random initial points. More details on the training mechanism can be found in \cite{ref13}.

\vspace{0.05cm}

\subsubsection{Event Scoring}
After training the basic model, the blocks of micro-PMU data stream are passed to the discriminator and the output is a scalar number which is defined as \emph{score}. We pass the whole training set to the discriminator and calculate the scores. A normal probability distribution function (pdf) is fitted to the obtained scores, i.e., $scores \sim \mathcal{N}(\mu,\sigma^{2})$, due to the fact that these scores must be very close to the global optimum, see (\ref{eq2}) and (\ref{eq3}). This is because of the infrequent nature of the events in power distribution systems. 

\vspace{0.05cm}

\subsubsection{Algorithm}

The proposed basic event detection method is summarized in Algorithm 1. It works based on the fact that events in micro-PMU data are infrequent. In fact, our analysis of the real-world micro-PMU data shows that events occur at {about} $0.04\%$ of the times. Thus, the default for the trained model must be the normal operation of the power distribution system. As a result, the discriminator is essentially trained to distinguish between the absence and the presence of the events, which is exactly what is needed in order to detect the events. 

%
%
It should be noted that, a common choice for $z_p$  in the threshold $\mu \pm z_p \sigma$ is 3, known as the three-sigma rule \cite{threesigma}. 


\subsection{Enhanced Method} \label{proposed GAN}

The basic method in Section II.A requires training a \emph{single} GAN model, where the features are $V$, $I$, $P$, and $Q$. However, given the characteristics of the micro-PMU data, in this section, we propose to develop and train \emph{two} separate GAN models, one for the voltage measurements $V$, and another one for the rest of the measurements, i.e., $I$, $P$, and $Q$. 


\vspace{0.05cm}


\subsubsection{Feature Analysis}

\begin{figure}[t]
\vspace{-0.3cm}
\captionsetup{font=footnotesize}
{\scalebox{0.31}{\includegraphics{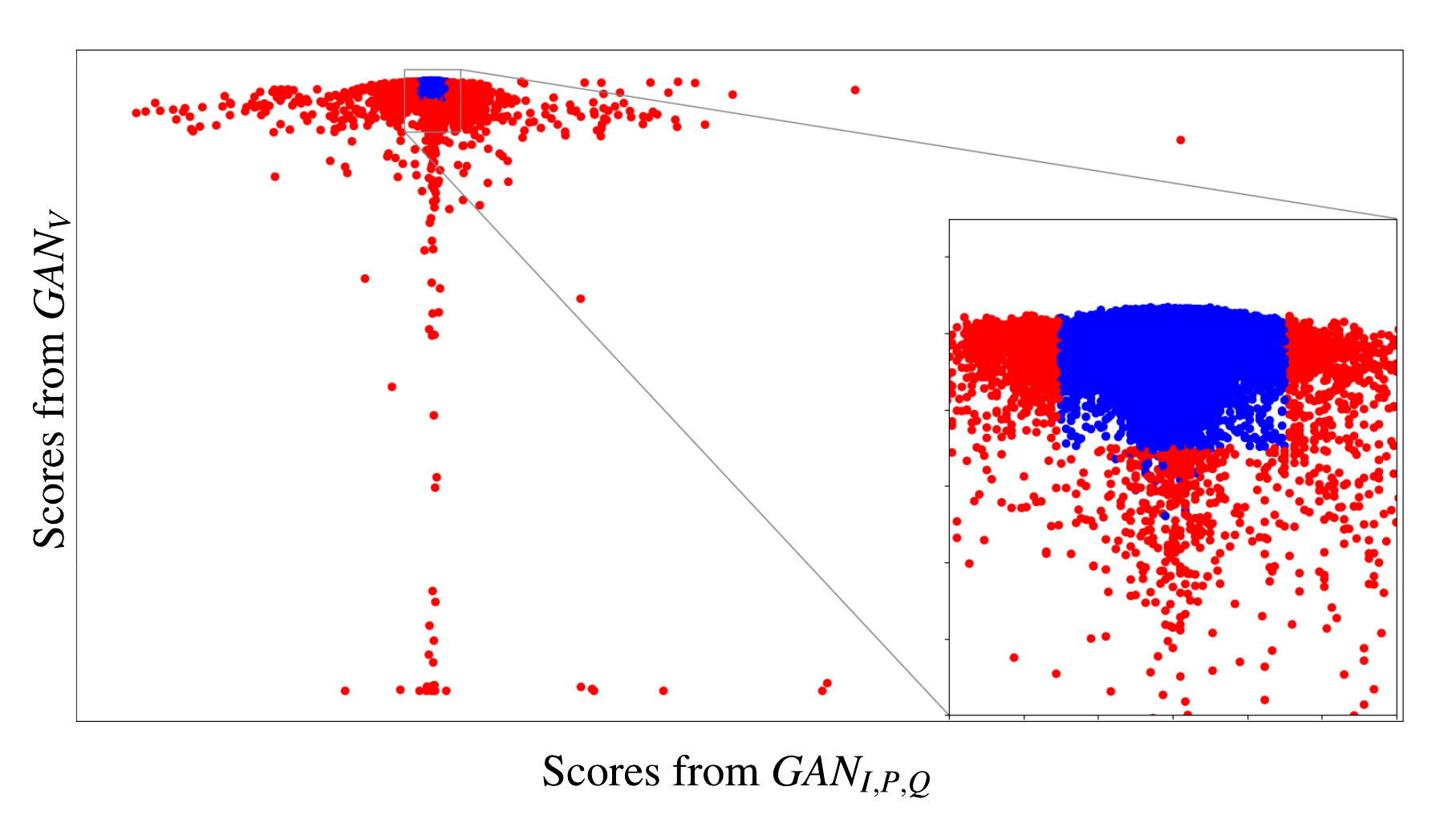}}}
\vspace{-0.5cm} \caption{The importance of using two GAN models in the enhanced method: while the scores from the $GAN_{I,P,Q}$ model can detect most events; there are events that are detected only if the scores from the $GAN_{V}$ model are also considered. Blue dots denote normal data while red dots denote events.}
\vspace{-0.3cm} \label{GAN_GANV_scores}
\end{figure}

After applying the  basic method to real-world micro-PMU data, we observed that Algorithm 1 sometimes fails to detect events that demonstrate signatures only in voltage magnitude. Such event cannot trigger the score to exceed the threshold. Further investigation revealed that this is because, in power distribution systems, voltage measurements are much less volatile than current measurements. Therefore, the GAN model sometimes cannot properly extract the characteristics of the voltage measurements. 



\vspace{0.05cm}

\subsubsection{Training Multiple GAN Models}

To remedy the above issue, we propose to construct two separate GAN models that are trained in parallel. One GAN model, denoted by $GAN_V$, has 3 features as its input, which are the voltage {magnitude} measurements across the three phases. The other GAN model, denoted by $GAN_{I,P,Q}$, has 9 features as its input, which are  current {magnitude}, active power, and reactive power measurements across the three phases. Importantly, it is observed that $I$ has high correlations with $P$ and even $Q$, which makes it desirable to combine $I$, $P$, and $Q$ into one GAN model; as opposed to having four GAN models for $V$, $I$, $P$, and $Q$.   


\vspace{0.05cm}

\subsubsection{Event Scoring}
Once each of the two GAN models is trained, the resulting Discriminator function is used to generate its own scores. An example for the scores that are generated by the two GAN models are shown in Fig. 1. The \emph{blue dots} represent \emph{normal data}. The red dots represent \emph{events}. We can see that each of the two GAN models detects only a sub-set of events. The events that are scattered across x-axis are the ones that are detected by $GAN_{I, P, Q}$. They include the majority of the events. The events that are scattered across y-axis are the ones that are detected by $GAN_{V}$. Thus, both GAN models are both needed to enhance accuracy of event detection. 


\vspace{0.05cm}

\subsubsection{Algorithm} 

The proposed enhanced event detection method is summarized in Algorithm 2. It works by examining the scores of the two separate GANs; thus having a dedicated deep learning architecture to detect the events in voltage magnitude and another deep learning architecture to detect the events that involve the current, active power, and reactive power. The rest of the algorithm is similar to Algorithm 1.



\begin{algorithm}[t]
    \caption{Event Detection - Enhanced Method}
  \begin{algorithmic}[t]
  
      \STATE \textbf{Input:} Training data and test data: $V$, $I$, $P$ and $Q$. 
    
      \STATE \textbf{Output:} Event Detection Flag $F$.

      \STATE \textbf{// Learning Phase}

    \STATE Train the $GAN_{I,P,Q}$ model.
    
    \STATE Use the Discriminator as scoring function $D_{I,P,Q}^*(\cdot)$.

    \STATE Calculate the scores for the training data.

    \STATE Fit a Normal PDF $\mathcal{N}(\mu,\sigma^{2})$ to the obtained scores. 

    \STATE Train the $GAN_{V}$ model.
    
    \STATE Use the Discriminator as scoring function $D_{V}^*(\cdot)$.

    \STATE Calculate the scores for the training data.

    \STATE Fit a Normal PDF $\mathcal{N}(\phi,\varphi^{2})$ to the obtained scores. 
    
    \STATE \textbf{// Detection Phase}
    
    \STATE \textbf{For} each new micro-PMU test data \textbf{Do}
    
    \STATE $ \ \ \ $ Calculate the score $s_1$ using $D_{I,P,Q}^*(\cdot)$.

    \STATE $ \ \ \ $ Calculate the score $s_2$ using  $D_{V}^*(\cdot)$.
    
    \STATE $ \ \ \ $ \textbf{If} $s_1 \notin (\mu - z_p \delta, \mu + z_p \delta)$ \textbf{or} 
    
    \STATE $ \ \ \ \ \ \ \ s_2 \notin (\phi - z_p \varphi, \phi + z_p \varphi)$ \textbf{Then}
    
    \STATE $ \ \ \ \ \ \ $ $F = 1$ // Event
    
    \STATE $ \ \ \ $ \textbf{Else} 
    
    \STATE $ \ \ \ \ \ \ $ $F = 0$ // No Event
    
    \STATE $ \ \ \ $ \textbf{End} 
    
    \STATE \textbf{End}

    \label{alg_event_scoring_enhanced}
  \end{algorithmic}
\end{algorithm}

\section{Experimental Results} \label{Results}


The proposed event detection methods are applied to the \emph{real-world} data from a 
distribution feeder in Riverside, CA \cite{ref9}. The resolution of the data is 120 readings per second. In total, 1.8 billion measurement points are analyzed. In particular, two weeks of data are used to train the GAN models. One day of data is used to test the event detection methods. 
%
%
%
Event detection is applied on windows of size 40 data points. Each window has an overlap of size 20 data points with the next window in order to assure not missing any event. 


\subsection{Performance Comparison}

The effectiveness of the event detection methods is investigated over 1000 reference events in micro-PMU data, that are visually extracted 
within a specific period of time.

The summary of the results are shown in Table I. We can see that the basic method significantly outperforms the benchmark statistical event detection method in \cite{ref9}. Furthermore, the enhanced method considerably outperforms the basic method. Next, we explain the underlying causes for these differences by going through several examples of the events that are detected. 

\begin{table}[t]
\begin{center}
\captionsetup{font=footnotesize}
\caption{Event Detection Accuracy}
\begin{tabular}{| l || c | c | c |}
\hline  & Benchmark \cite{ref9} & Basic Method & Enhanced Method \\
\hline Accuracy & 0.3640 & 0.6943 & 0.8805 \\
\hline F1-score & 0.3614 & 0.7676 & 0.9023 \\
\hline
\end{tabular}
\end{center}
\label{t1} \vspace{-0.5cm}
\end{table}

\subsection{Assessment of the Basic Method} \label{GAN}

\begin{figure}[h!]
\vspace{-0.2cm}
\captionsetup{font=footnotesize}
\hspace{-5mm}
{\scalebox{0.2}{\includegraphics{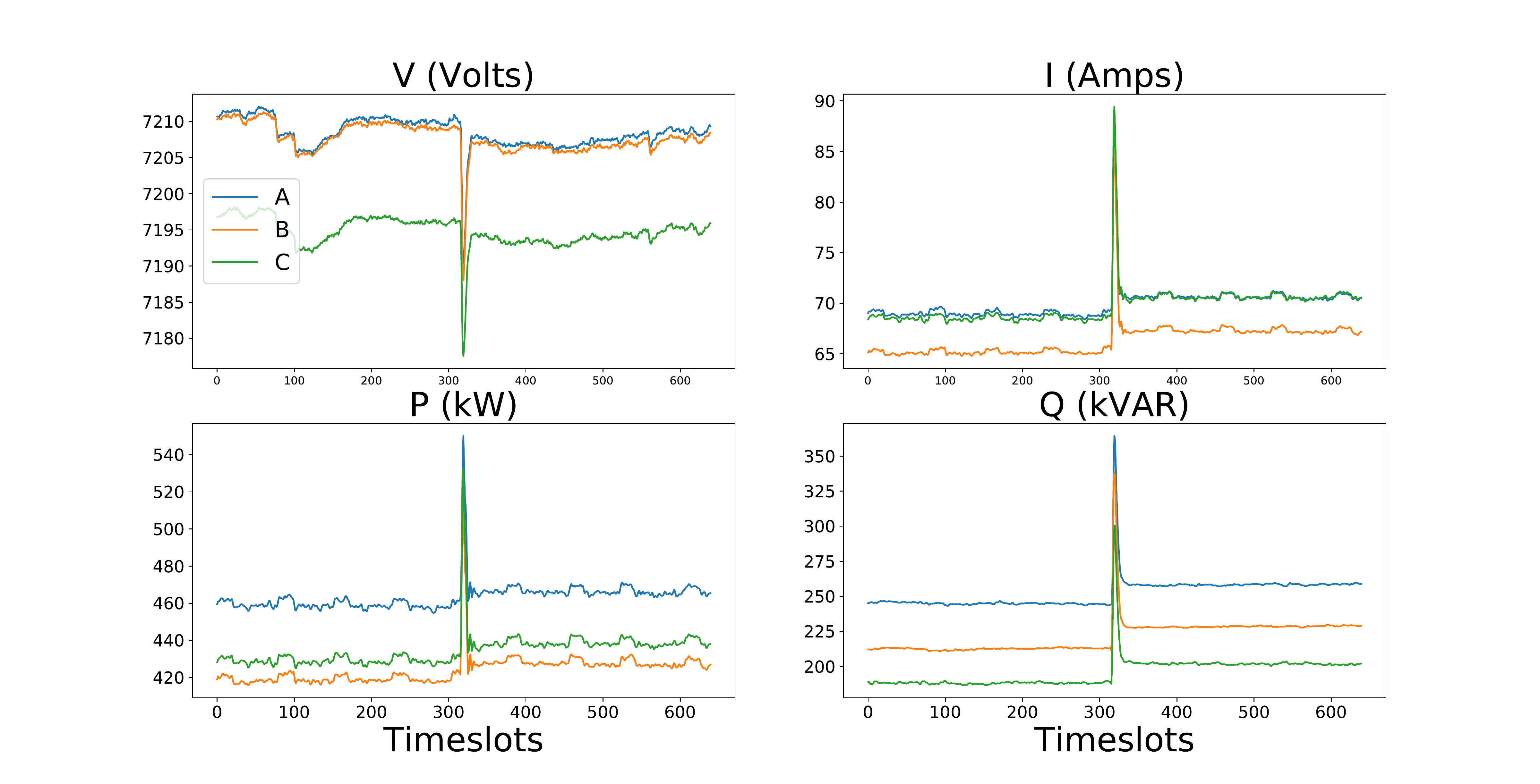}}}
\vspace{-0.5cm} \caption{Inrush current  with impact on all features. This event is detected by all the three methods: statistical, basic, and enhanced.}
\vspace{-0.1cm} \label{GANS_peak}
\end{figure}
\begin{figure}[h!]
\vspace{-0.35cm}
\captionsetup{font=footnotesize}
\hspace{-5mm}
{\scalebox{0.2}{\includegraphics{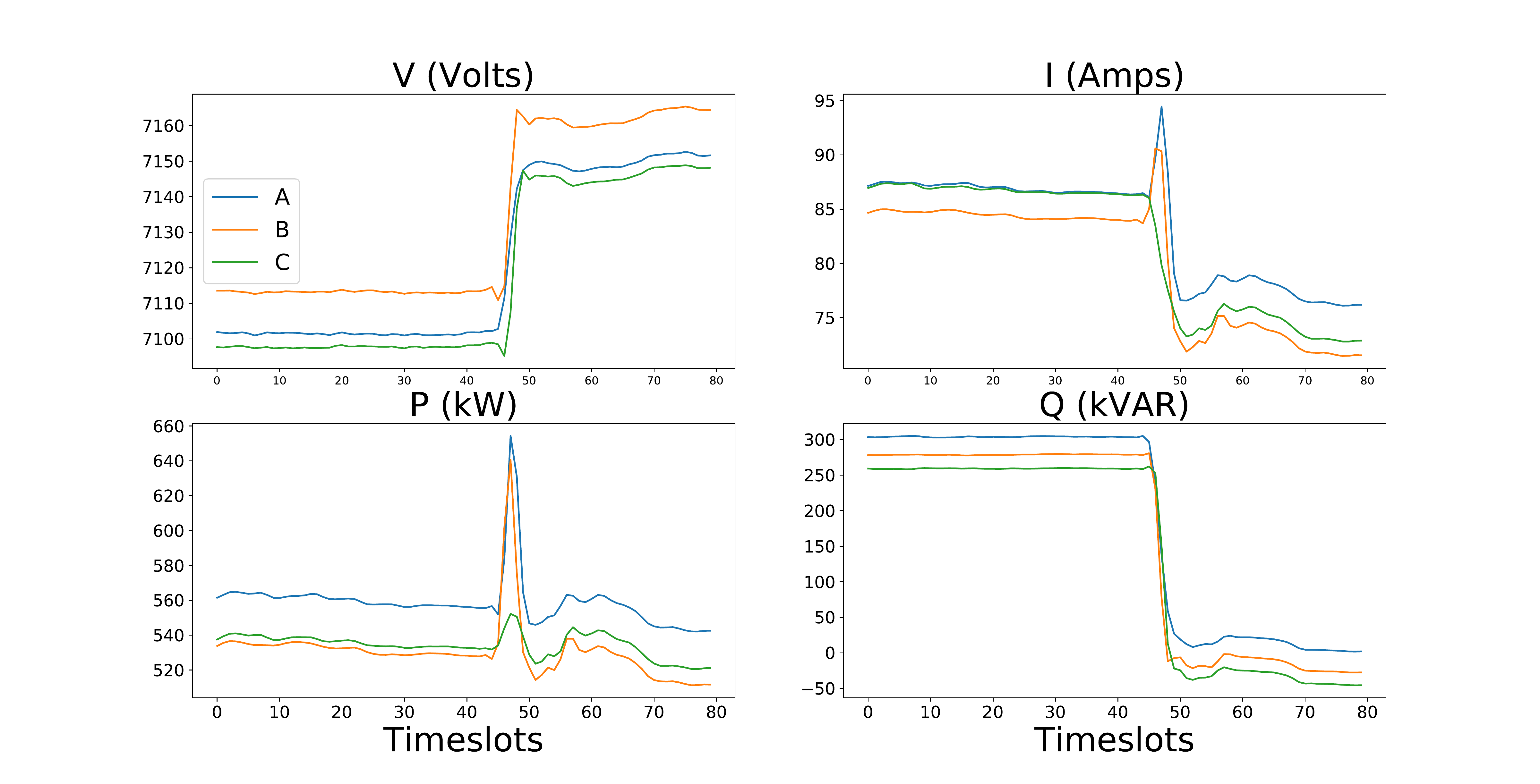}}}
\vspace{-0.6cm} \caption{Capacitor bank switching with impact on all features. This event is detected by all the three methods: statistical, basic, and enhanced.}
\vspace{-0.3cm} \label{GANS_cap}
\end{figure}
\begin{figure}[h!]
\captionsetup{font=footnotesize}
\hspace{-5mm}
{\scalebox{0.2}{\includegraphics{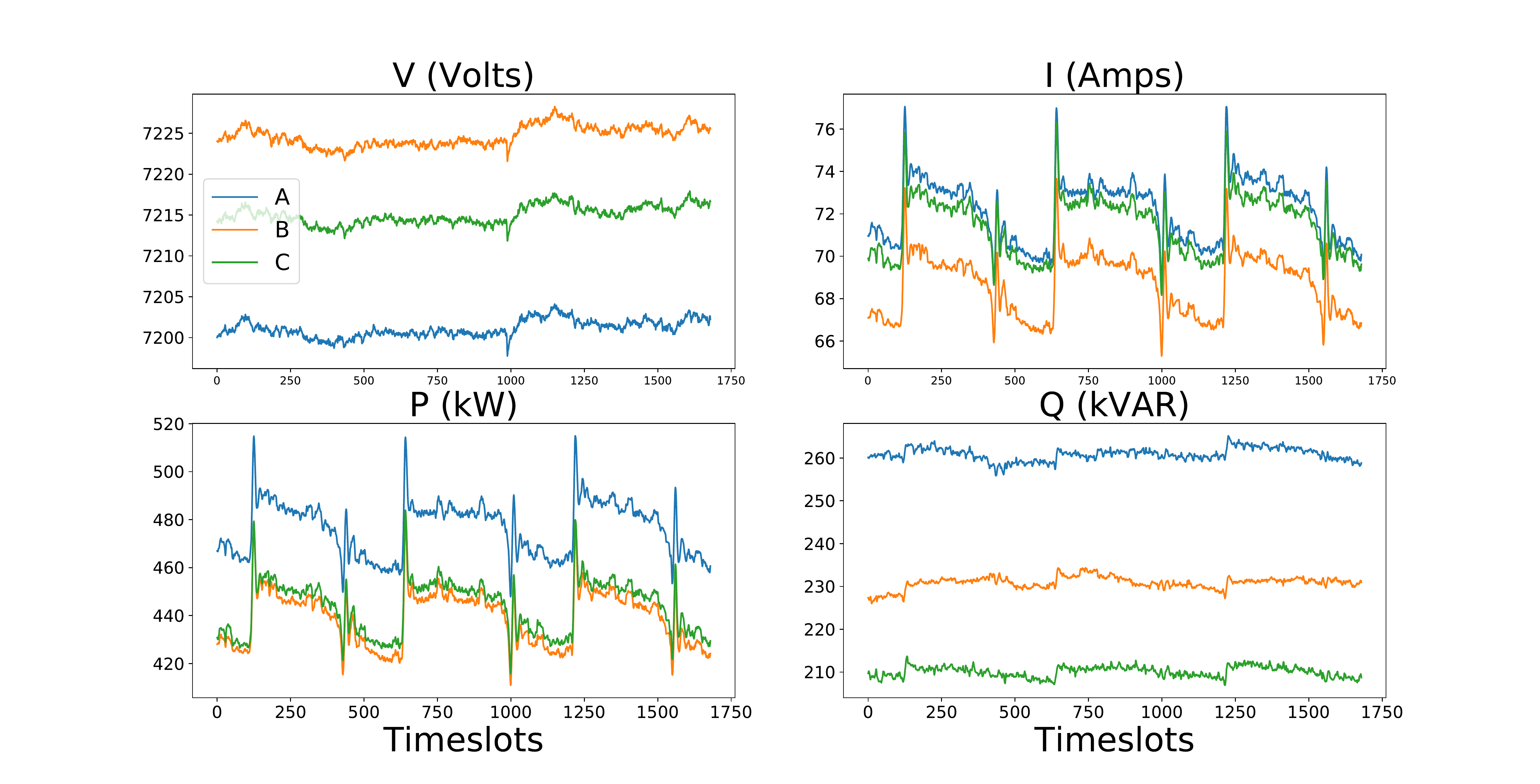}}}
\vspace{-0.6cm} \caption{An event with major impact only on current and active power. This event is detected by the basic method, but not by the statistical method.}
\vspace{-0.25cm} \label{GAN_just_I_P}
\end{figure}
\begin{figure}[h!]
\captionsetup{font=footnotesize}
\hspace{-5mm}
{\scalebox{0.2}{\includegraphics{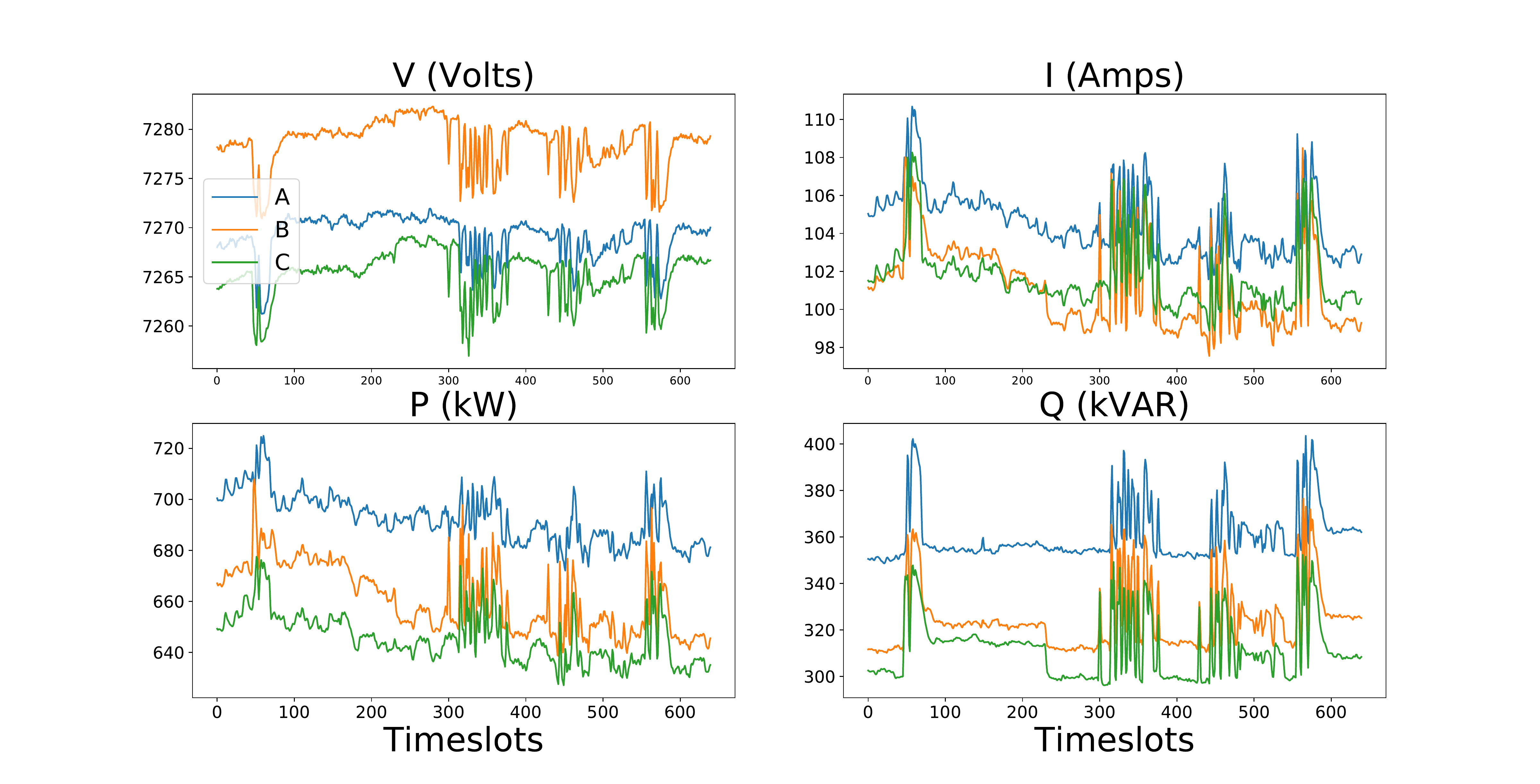}}}
\vspace{-0.6cm} \caption{An event involving oscillations. This event is detected by the basic method, but it is not detected by the statistical method.}
\vspace{-0.25cm} \label{high_freq_all_separate}
\end{figure}
\begin{figure}[h!]
\captionsetup{font=footnotesize}
\hspace{-5mm}
{\scalebox{0.2}{\includegraphics{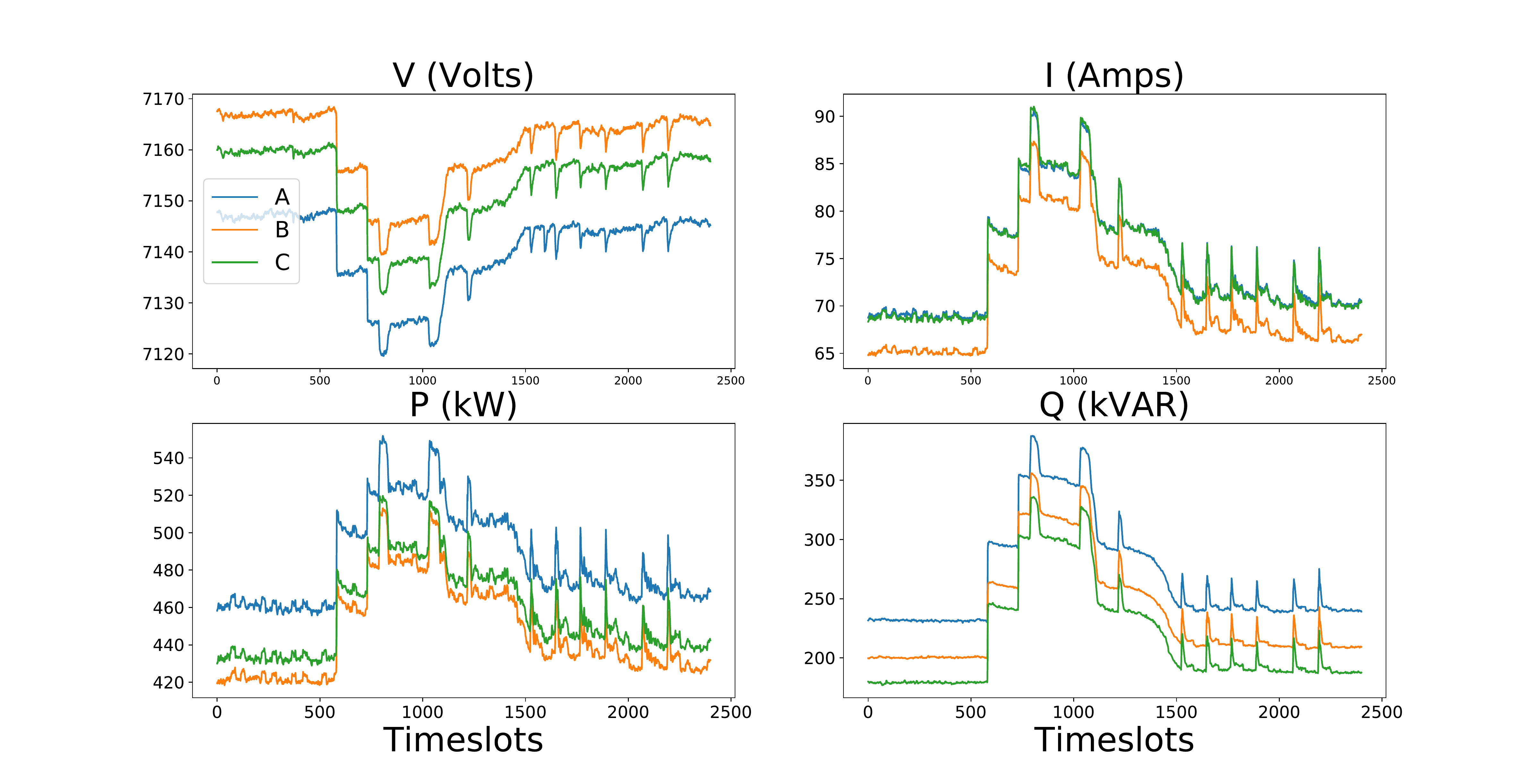}}}
\vspace{-0.6cm} \caption{A rare and long event with 20 seconds of transient signature. All pieces of this long event are detected and captured by the basic method. The statistical method only detects a step change at the beginning of this event.}
\vspace{-0.35cm} \label{GANS_20sec}
\end{figure}

Figs. \ref{GANS_peak} to \ref{GANS_20sec} show five examples of the events that are detected by the basic method. Importantly, the prevalent statistical method in \cite{ref9} detected only the first two of such events. Regarding the events in Figs. \ref{GAN_just_I_P} and \ref{high_freq_all_separate}, they are not detected by the method in \cite{ref9} because the changes in the magnitudes are relatively small and do not significantly affect the statistical measures, such as the absolute deviation around median. As for the event in Fig. \ref{GANS_20sec}, all the pieces of this long event are detected by the basic method at several subsequent windows of the data. However, the statistical method in \cite{ref9} only captures the step change the beginning of this event; because the statistical characteristics remain the same afterwards. 


%

\subsection{Assessment of the Enhanced Method} \label{proposed}

Figures 7 and 8 show two events that are detected by the enhanced method. But they are not detected by either the prevalent statistical method in \cite{ref9} or even the basic method. The basic method fails to detect these two events because the main signatures are in voltage and they are relatively small in magnitude. Therefore, only the additional GAN model in the enhanced method can capture these events. This demonstrates the importance of the change in the model that was proposed in the enhanced method. Regarding the event in Fig. 8, it demonstrates momentary oscillations that started only after some sort of actions, possibly a tap changing event, where the oscillations damped after a short period of time. Events like this are important, for example, for asset monitoring. However, only the enhanced method was able to detect such event.

\begin{figure}[t]
\vspace{-0.1cm}
\captionsetup{font=footnotesize}
\hspace{-5mm}
{\scalebox{0.2}{\includegraphics{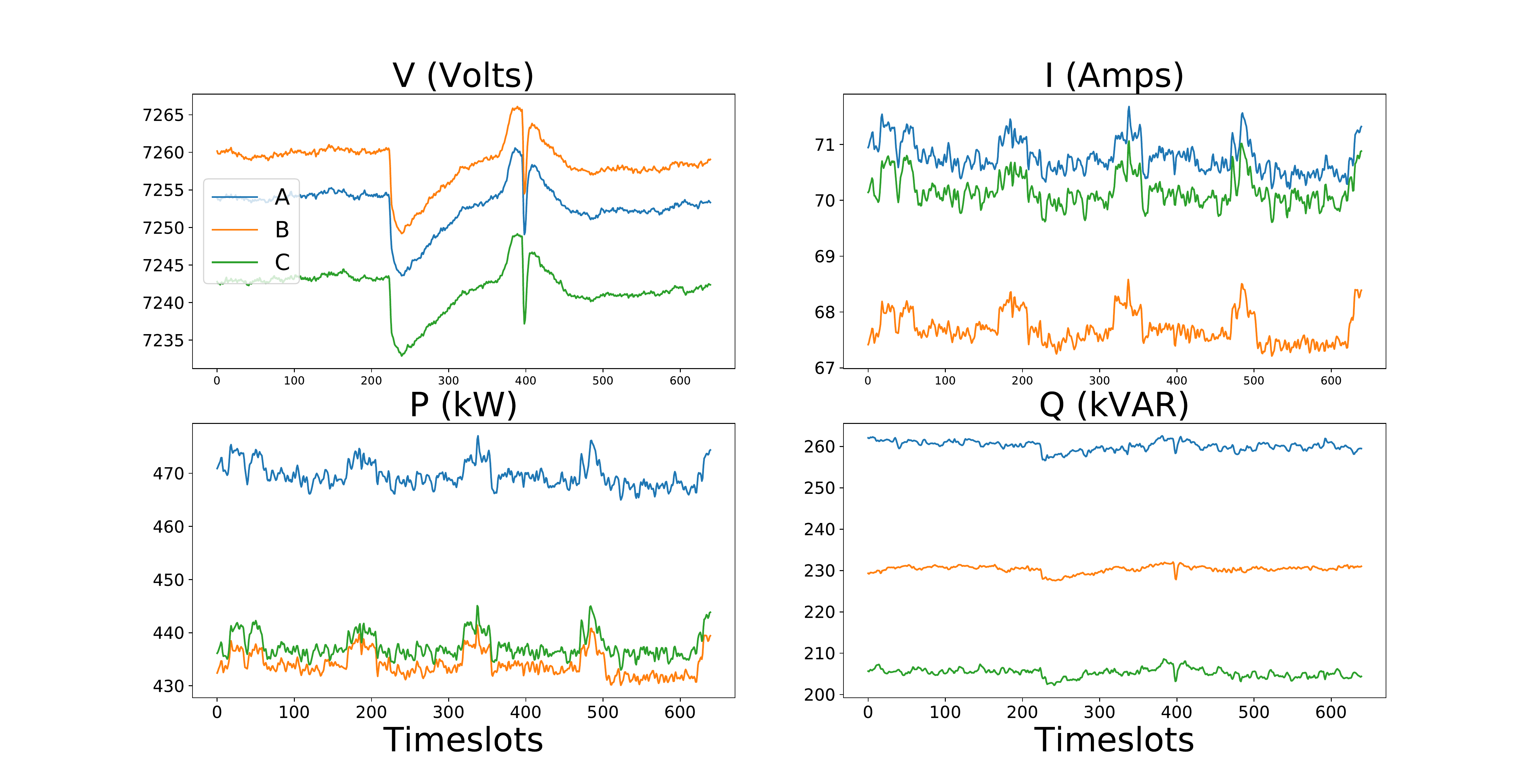}}}
\vspace{-0.55cm} \caption{An event with impact mainly on voltage. It is detected by the enhanced method. But it is not detected by the basic method or the statistical method.}
\vspace{-0.5cm} \label{GANV_two_drop}
\end{figure}

\begin{figure}[t]
\vspace{0.2cm}
\begin{center}
\captionsetup{font=footnotesize}
\hspace{20mm}
{\scalebox{0.294}{\includegraphics{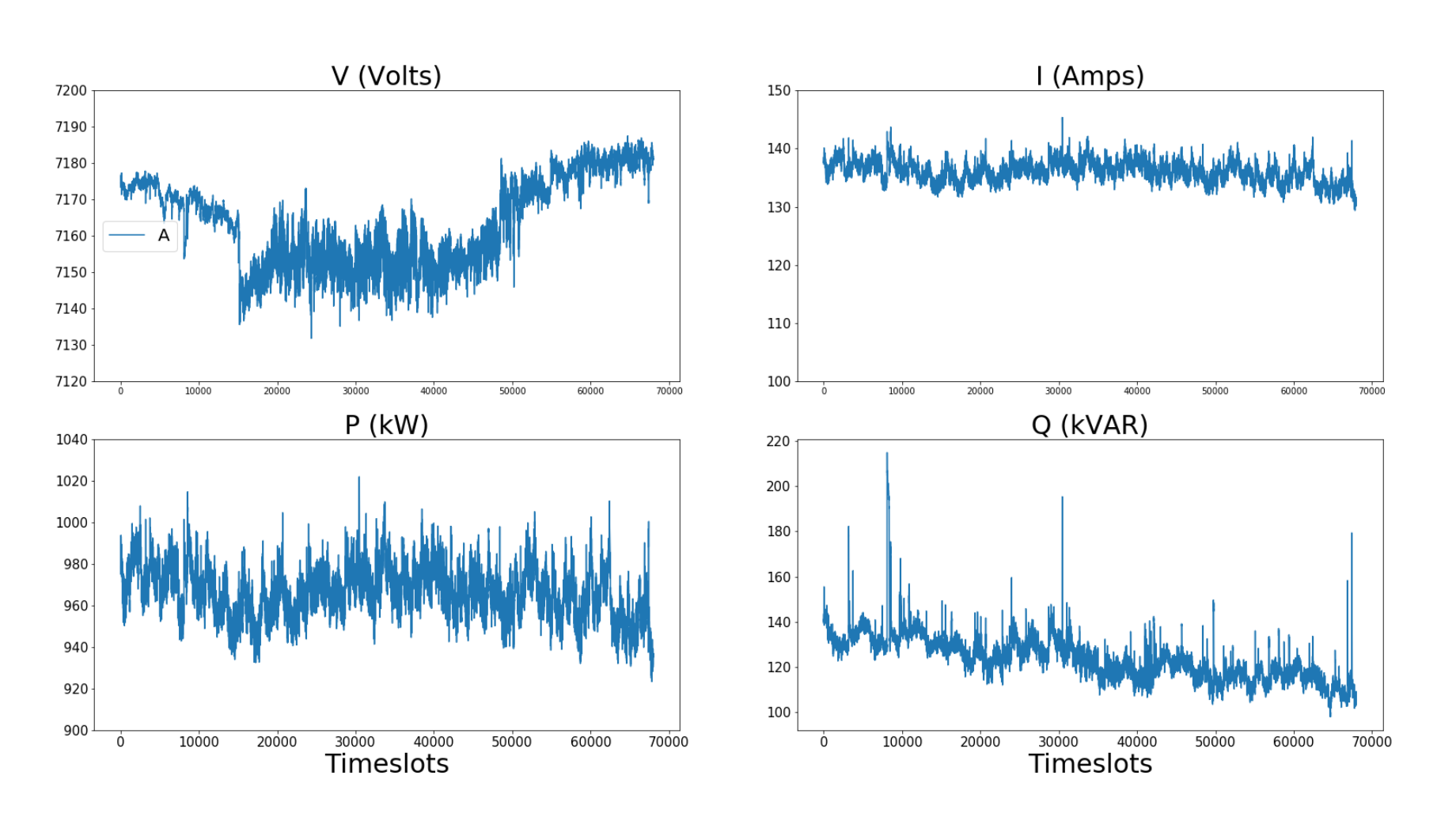}}}
\vspace{-0.2cm}
\caption{An event with momentary and damping oscillations in voltage, shown on one phase only. This event is detected by the enhanced method. But it is not detected by the basic method or the statistical method.}
\end{center}
\vspace{-0.5cm} \label{high_freq_all}
\end{figure}

\section{Conclusions}
\label{conslusion}
Two novel unsupervised deep learning methods are proposed to detect events in micro-PMU data streams. They work by constructing  Generative Adversarial Network (GAN) models. They are capable of extracting the characteristics of a wide verity of events in large volumes of micro-PMU data. The basic method involves a single GAN model. The enhanced method is equipped with additional analysis of features. It involves training two parallel GAN models. Both methods are capable of {detecting events with} point-signatures and group-signatures. They are particularly well-suited to detect the events in distribution systems where the event may impact only a subset of the features and only or two phases; in addition to the cases that all three phases are affected. Real-world data from micro-PMU field installation is used to evaluate the performance of the proposed event detection methods. It is observed that the basic method significantly outperforms a prevalent statistical event detection method in the literature. Furthermore, the enhanced method considerably improves the performance over the basic method. Several examples of the events that detected by different methods are shown and discussed in order to understand the characteristics of the proposed unsupervised event detection methods. 



\bibliographystyle{IEEEtran}
\bibliography{main}

\end{document}